**Decoding Positive Selection in *Mycobacterium tuberculosis* with Phylogeny-Guided Graph Attention Models**

Linfeng Wang[1], Susana Campino[1], Taane G. Clark[1,2,*], Jody E. Phelan[1,*]

[1] Faculty of Infectious and Tropical Diseases, London School of Hygiene & Tropical Medicine, WC1E 7HT London, UK

[2] Faculty of Epidemiology and Population Health, London School of Hygiene & Tropical Medicine, WC1E 7HT London, UK

* Joint corresponding authors

Jody.phelan@lshtm.ac.uk, taane.clark@lshtm,ac.uk

Department of Infection Biology,

Faculty of Infectious and Tropical Diseases

London School of Hygiene & Tropical Medicine, Keppel Street, London, UK

**Abstract**

Positive selection is a key evolutionary force in *Mycobacterium tuberculosis*, driving the emergence of adaptive mutations that influence drug resistance, transmissibility, and virulence. Phylogenetic trees capture the hierarchical evolutionary relationships among isolates, making them an ideal framework for detecting such adaptive signals. Here, we present a phylogeny-guided graph attention network approach, coupled with a novel method for converting SNP-annotated phylogenetic trees into graph structures suitable for graph neural network processing. Using a dataset of 500 *M. tuberculosis* isolates, representing the four main lineages, and 249 single-nucleotide variants (84 resistance-associated and 165 neutral) spanning 61 drug-resistance genes, we constructed graphs where nodes represented individual isolates and edges reflected phylogenetic distances. To reduce noise and highlight local evolutionary structure, we pruned edges between isolates separated by more than seven internal nodes. Node features were encoded as binary indicators of SNP presence or absence, and the graph attention network (GAT) architecture comprised two attention layers with a residual connection, followed by global attention pooling and a multilayer perceptron classifier. The model achieved an accuracy of 0.88 on the held-out test set, and application to 146 WHO-classified "uncertain" variants identified 41 candidates with convergent occurrence across multiple lineages, consistent with adaptive evolution. These variants included: *eis* c.-37G>T (kanamycin, amikacin), *embA* c.-12C>T (ethambutol), *rpoA* Thr187Ala (rifampicin), and *rpoC* Leu516Pro (rifampicin). These findings demonstrate both the feasibility of transforming phylogenetic trees into graph neural network-compatible structures and the utility of attention-based models (e.g., GATs) for detecting signals of positive selection, supporting genomic surveillance and prioritising candidate variants for experimental validation.

**Introduction**

Tuberculosis (TB), caused by *Mycobacterium tuberculosis*, remains one of the leading causes of infectious disease mortality worldwide, responsible for 1.09 million deaths in 2024[1]. The development of drug resistance has exacerbated this epidemic, with 400,000 of all new TB cases reported to have resistance to at least the first-line drug rifampicin. A key driving force underpinning such an increase in the incidence of drug-resistant TB spread is the emergence of mutations in drug targets and pro-drug activators. In the presence of a drug, a selective pressure is exerted on the bacteria, which causes resistance-conferring mutations to rapidly reach fixation in the population. *M. tuberculosis* has a genome of approximately 4.4 Mb[2], with an estimated mutation rate of between $2\times 10^{-10}$ and $3\times 10^{-10}$ substitutions per site per generation, equivalent to approximately 0.2–0.5 single-nucleotide polymorphisms (SNPs) per genome per year[3]. The evolution of *M. tuberculosis* is driven by the selection of random mutations that confer a fitness advantage, thereby increasing their frequency in the population. These positively selected mutations may reflect evolutionary responses to drug pressure, host immunity, or transmission dynamics. Importantly, the evolutionary trajectory of *M. tuberculosis* is shaped by its strictly clonal population structure, with no recombination. This allows for the reconstruction of its global phylogeny into distinct lineages[4], each representing deep evolutionary splits. These lineages not only mirror the geographic and demographic history of TB spread, but also capture lineage-specific patterns of drug resistance, virulence, and transmissibility.

Phylogenetic trees are graphical representations of the evolutionary relationships among a set of organisms or genetic sequences, inferred from genetic, genomic, or phenotypic data[5]. In the context of infectious diseases such as TB, phylogenies help to reconstruct the ancestral relationships between *M. tuberculosis* isolates, often reflecting historical transmission events and (lineage-based) population structure[6]. These trees are typically rooted, bifurcating structures in which branch lengths may reflect genetic divergence or time elapsed since a common ancestor. This structure provides a natural framework for studying evolutionary signals, as it explicitly captures the hierarchical ancestry between samples. When mutations are mapped onto these trees, the topology can reveal patterns consistent with positive selection - such as the repeated emergence of mutations across distinct lineages, clustering on terminal branches, or association with rapidly expanding clades. These patterns suggest that certain positively selected mutations confer a fitness advantage, such as increased transmissibility or resistance to drugs. Notably, such mutations may arise independently in different lineages but are consistently retained, reflecting strong and recurrent adaptive pressures acting on the pathogen population. Identifying these mutations offers an opportunity to uncover the genetic basis of adaptation, revealing both known and previously uncharacterised functional loci that contribute to pathogen persistence and spread.

Graph neural networks (GNNs) are a class of models designed to operate directly on structured graph data. By representing phylogenetic trees as computational graphs, GNNs are uniquely positioned to incorporate both local mutational features and the global topology of evolutionary relationships. Through iterative message passing, these models can learn context-dependent representations that reflect how mutations are embedded within their phylogenetic landscape[7]. A graphical convolutional neural network (GCN) is a classical formulation of the GNN architecture, in which each node aggregates features from its connected neighbours and passes this aggregated signal through neural layers to extract localised information. However, standard GCNs assign uniform or fixed weights to neighbouring nodes, potentially underutilising informative signals in uneven or heterogeneous graph structures. To address this limitation, Graph Attention Networks (GATs)[8] introduce an attention mechanism that allows the model to learn the relative importance of each neighbouring node during message passing. In this framework, attention coefficients are computed dynamically and used to weight each neighbour's contribution, enabling the network to focus more strongly on structurally or mutationally relevant regions of the phylogeny. This selective aggregation is particularly advantageous in evolutionary graphs, where certain branches or recurrent mutations may carry disproportionately informative signals of adaptation.

Here, we investigate the application of GNNs to the task of detecting positively selected variants in *M. tuberculosis*, using SNP-annotated phylogenetic trees as input. Our approach encodes both tree structure and mutational data into a unified graph model, enabling the classifier to distinguish adaptive mutations and those likely to arise under neutral evolution and be passed on due to simple transmission.

**Methods**

*Data structure design and transformation*

To evaluate the capability of graph neural networks in processing phylogenetic tree-structured data, we implemented a GAT model on SNP-annotated *M. tuberculosis* phylogenies. This proof-of-concept aims to assess whether GATs can effectively leverage phylogenetic context to identify functionally relevant SNP patterns across related samples. As illustrated in the simulated tree (**Figure 1a**), a SNP appears in samples B, D, and G. Pairwise phylogenetic distances are calculated by counting the number of internal nodes separating each sample (**Figure 1b**). In this illustrative example, edges between samples separated by more than four internal nodes were removed to highlight local phylogenetic structure. For real datasets, an optimised cutoff was selected empirically to balance graph sparsity and information retention. This sparsification step was crucial to negate any noisy long-range interactions and emphasise meaningful local neighbourhoods within the tree topology. The resulting subgraph (highlighted in blue in **Figure 1b**) preserves a clear dual-cluster structure, which is also reflected in the corresponding graph visualisation (**Figure 1c**). The updated representation of node G ($H'_g$) is obtained by combining the attention-weighted features of its connected neighbours (**Figure 1c**). Each node in the graph was initialised with a binary feature vector indicating the presence or absence of the target SNP, this is the summed with an edge-aware bias.

In a more generalised equation (**Figure 1d**), for a target node $i$, the updated embedding $h'_i$ is computed by aggregating messages from its neighbours $j$. The attention score is calculated using a learnable weight vector $\alpha \in \mathbb{R}^{F'}$, applied to the concatenated transformed features of the target node $h_i$ and its neighbor $h_j$, each projected via a shared linear transformation $W \in \mathbb{R}^{F'+F}$. The concatenated representation $[Wh_i \parallel Wh_j]$ denotes node-pair interactions. Edge information is integrated via a scalar product between a learnable bias parameter $b \in \mathbb{R}$ and the edge length, $edge_{ij}$, representing the phylogenetic distance (internal node count) between nodes $i$ and $j$. The full attention coefficient is then passed through a non-linear activation function $\sigma$, here LeakyReLU, to introduce non-linearity before SoftMax normalisation and attention-weighted message aggregation.

*Data sources*

The genomic data were derived from transmission records of a multidrug-resistant (MDR+) enriched subset of the in-house 100K dataset, collected from publicly available sources[9]. For each SNP, a rooted phylogenetic tree was first reconstructed from 500 *M. tuberculosis* clinical samples, comprising the four main lineages (L1-L4). These phylogenies were then transformed into graphs, where nodes represented individual genome samples and edges reflected undirected phylogenetic connections. Node features were encoded as a single binary variable denoting the presence or absence of the focal SNP. In total, 249 SNP-specific graphs were generated, each corresponding to one phylogenetic tree, with labels assigned according to WHO catalogue classifications of drug resistance (https://github.com/GTB-tbsequencing/mutation-catalogue-2023/tree/main/Final%20Result%20Files). To implement the model, we applied it to a set of 146 mutations currently classified by the WHO as having "uncertain" associations with drug resistance. Graphs were constructed using PyTorch Geometric[10] and batched using its native DataLoader. The visualisation of the phylogenetic tree was performed using networkX[11] and iTOL[12] tools.

*Bioinformatics pipeline*

The raw sequence reads were trimmed using trimmomatic[13] according to the sequence quality generated using fastQC[14]. The trimmed reads were then mapped to the H37Rv reference genome using BWA-mem[15]. SNPs were called using the BCF/VCF[16] tool suite on regions with >10-fold read depth coverage. SNPs were converted into a FASTA format alignment, which was used as input to RAxML[17] to reconstruct the phylogeny. All drug resistance mutations were excluded when building the tree to prevent data leakage.

*GNN architecture and training*

To classify SNP presence using phylogenetically structured data, we employed Bayesian hyperparameter optimisation to identify the optimal configuration for model training. The final architecture was implemented using PyTorch Geometric (v2.4.0)[10] and consisted of a two-layer Graph Attention Network (GAT) as shown in **Supplementary Table 2**. Edge-level features (phylogenetic edge lengths) were used as scalar edge length during message passing and were supplied to both GATConv layers via the edge_attr argument. The first GATConv layer used 8 attention heads, each producing 32 output features, resulting in a concatenated 128-dimensional output. This was followed by batch normalisation, Exponential Linear Unit (ELU) activation, and dropout (p = 0.29). The second GATConv layer maintained the same dimensionality (128 features) with a single attention head and concat=False to allow residual connection with the first layer. The output of this layer was batch-normalised, passed through an ELU activation, and added to the first layer's output via a residual connection. Dropout was applied after both GAT layers. For graph-level prediction, node embeddings were aggregated using a global attention pooling layer with a linear gating function to compute attention scores across nodes. The resulting pooled feature vector was passed through a fully connected layer with 32 hidden units, followed by batch normalisation, rectified linear unit (ReLU) activation, and dropout. A final linear layer projected the representation to a two-dimensional output space for binary classification. Class probabilities were obtained by applying a SoftMax function over the two logits. The probability corresponding to the predicted class was then passed through a sigmoid transformation to yield a confidence score between 0 and 1. For negative predictions (probability < 0.5), the confidence was defined as $1 - p$, ensuring that the reported value consistently reflects the model's certainty in its assigned label.

Training was conducted using the Adam optimiser in conjunction with a one-cycle learning rate policy. The learning rate was annealed from a minimum of 9.89 × 10⁻⁵ to a maximum of 7.55 × 10⁻⁴ using a cosine annealing schedule. The loss function used was cross-entropy with class weights of 1.0 for negative cases and 1.17 for positive cases to address mild class imbalance. The model was trained for a maximum of 600 epochs, with early stopping applied based on validation loss plateauing. The 249 mutation tree data were split into training (149), validation (50), and a held-out test set (50).

*Edge attention*

Attention coefficients were obtained from the GAT using the PyTorch Geometric GATConv layers. Unless stated otherwise, attention was extracted from the first GAT layer. Multi-head coefficients were averaged across heads to yield a single edge-level score. Attention scores were aligned to the undirected NetworkX graph used for visualisation by matching both (u,v) and (v,u) tuples. Undirected graphs were rendered with NetworkX using a spring layout. Nodes were coloured by mutation presence, and edges were coloured by attention weights.

To quantify how concentrated the model's focus is, the Top-k Attention Mass (TAM) was computed, $TAM(k) = \frac{\sum_{i=1}^{m} \alpha_{\pi(i)}}{\sum_{e=1}^{E} \alpha_e}$, where $\alpha = \{\alpha_e\}_{e=1}^{E}$ denotes the averaged attention scores for the E edges. Sorting in descending order gives indices π such that $\alpha_{\pi(1)} \geq \alpha_{\pi(2)} \geq ¼$. For a chosen fraction $k \in (0,1]$, $m = kE$ edges are retained.

**Results**

*GAT training with supervised learning*

The dataset used in this study comprises 500 *M. tuberculosis* samples, spanning the four main lineages (L1 8, L2 175, L3 109, L4 223, mixed infections 15), and 249 SNP variants identified in 61 drug resistance genes. Among these, 84 mutations (33.7%) were annotated by the WHO as being associated with drug resistance, while 165 were classified as not associated with resistance. A total of 249 graph instances were generated, each corresponding to a single variant. Each graph shared a common underlying topology derived from a simplified phylogenetic tree, with node labels indicating the presence or absence of the specific mutation under consideration.

Using the data structure design described in **Figure 1**, the *M. tuberculosis* dataset (tree number=249, node number =500) was sparsified by removing edges between sample pairs separated by more than seven internal nodes in the phylogenetic tree (cut-off: edge length > 7). This sparsification step significantly simplified the resulting graph and reduced the number of fully connected nodes to four. By avoiding a fully connected architecture, the model was able to capture not just pairwise relationships but also the phylogenetic distances and the relative importance of connectivity between nodes. The effect of this cutoff was to prune the tree into a more interpretable network structure, revealing two distinct clusters connected by a small number of high-degree nodes acting as bridges (**Supplementary Figure S1**). This topology reflects localised phylogenetic neighbourhoods and supports the model's ability to learn from evolutionary context without being overwhelmed by excessive graph density and connectivity. The GAT model exhibited strong predictive performance when trained on this graph-encoded phylogenetic structure. On the test (holdout) dataset consisting of 50 variant graphs (17 positive selected, 33 neutral selection), the model achieved an accuracy of 0.88, AUC of 0.89 and an F1 score of 0.81. The resulting confusion matrix corresponds to a sensitivity of 0.76 and a specificity of 0.94, underscoring the model's capacity to distinguish positively selected mutations from neutral variants in a topology-aware manner (**Table 1**). Among the 50 variants in the test dataset, 29 were non-synonymous mutations (stop-gained 1, missense 28), 13 were synonymous mutations, 2 were annotated as non-coding transcript exon variants, and 6 were upstream gene variants. Of the mutations predicted by the model to be under positive selection, 10 were missense mutations, 1 was a stop-gained variant (*pncA* Tyr103*, pyrazinamide), 1 was a non-coding transcript exon variant, and only a single synonymous mutation was included (*embA* c.114a>9, ethambutol). Notably, synonymous mutations are not typically associated with functional changes or drug resistance and are therefore unlikely to be subject to positive selection. The near-complete exclusion of synonymous variants from the model's positively selected set supports its ability to prioritise biologically meaningful variation, underscoring the model's potential to capture signals of adaptive evolution relevant to drug resistance.

*Attention highlights mutation-diverse hubs*

As shown in **Figure 2**, attention is concentrated on edges connecting central nodes, forming a hub of high-weight connections. These edges predominantly link the central nodes to a large cluster of samples with diverse mutation profiles. By contrast, edges within the smaller, more homogeneous cluster receive markedly lower attention. This distribution is quantified by the TAM, where the highest-ranked 10% (k=10%) of edges capture 44.1% of the total attention. Such enrichment indicates that the model preferentially amplifies signals arising from mutation-diverse regions of the graph, which are more likely to carry predictive value for downstream classification.

*Testing on uncertain mutations*

To evaluate the model's utility beyond annotated resistance mutations, we tested it on 146 candidate variants classified as "uncertain" in the WHO catalogue. Of these, 27 (18.5%) were predicted by the model to be under positive selection. Previously found Putative drug resistance mutations and compensatory mutations were identifdied. Two putative mutations linked to drug resistance were identified: *rpoC* Leu483Ala (rifampicin) and *ubiA* Val188Ala(ethambutol). Additional variants in *ubiA* were also detected (Ala249Thr, Arg240Cys); this gene encodes a protein essential for cell wall synthesis and is implicated in ethambutol resistance. Furthermore, two mutations were observed in *Rv0010c* (c.-80A>G, c.-78A>G), which encodes a conserved membrane protein[18].

Similarly, variants were found in *whiB6* (c.-42G>T, Thr51Pro), a regulator of ESX-1 gene expression, a secretion system required for mycobacterial pathogenesis. Finally, we identified mutations of uncertain resistance significance, notably *whiB* Thr51Pro (present in 48.64% of MDR+ isolates in the 100k dataset, spanning 35 sublineages) and *rpoC* Ile491Val (49.69% of MDR+ isolates, spanning 44 sublineages) (**Supplementary Table S1**).

**Discussion**

TB remains a leading global health threat, fuelled by the emergence of drug-resistant strains. This study explores the application of Graph Attention Networks (GATs) to detect signals of positive selection in *M. tuberculosis* phylogenies, which can serve as genomic signatures of drug resistance. The model demonstrated promising performance, yielding balanced predictions despite being trained on a small and imbalanced dataset. A deeper examination of the attention weights in the edges revealed that the model was capable of selectively focusing on the most informative regions of the graph, while attenuating attention toward genetically uniform areas. This behaviour underscores the model's ability to prioritise predictive signals and is sufficient to establish a proof of concept. It also validates the use of internal node counts as edge lengths for converting phylogenies into graph neural network (GNN)-trainable data, an approach that simplifies the phylodynamic graph structure while preserving essential information, thereby supporting its novel application for the detection of positive selection. Importantly, the model does not explicitly test for dN/dS ratios or phylogenetic convergence, but instead learns patterns in graph topology, edge connectivity, and SNP distributions across isolates that distinguish resistance-associated from neutral variants.

The use of drug resistance as a proxy for positive selection may introduce limitations. Certain mutations may enhance bacterial survival or transmission without directly conferring drug resistance. For example, the *whi6* locus affects virulence by acting on ESX-1 gene expression, regulating virulence factor secretion. Additionally, a proportion of the reported uncertain mutations were previously found putative drug resistance mutation and compensatory mutations (**Table S1**). Therefore, the nature of the positive selection of these mutations may not only contribute to drug resistance but also to other mechanisms of improving the fitness of the bacteria.

While the model may correctly identify such variants, their mislabelling during training could lead to confusion, reduced training efficiency, and an upper limit on predictive performance.

Additionally, as with human interpretation of phylogenetic trees, the method's effectiveness depends on the number of truly positively selected mutations and the size of the dataset. In this context, our analysis of the uncertain section of the WHO catalogue provides an important demonstration of the model's broader utility. We identified a set of candidate mutations that may contribute to *M. tuberculosis* spread through drug resistance or enhanced bacterial fitness. Many of these exhibit convergent emergence in resistant lineages, underscoring their potential functional relevance (e*mbA c.*-43G>C, *ubiA* Ala249Thr, *Rv0010c* c.-80A>G, *whiB6* Thr51Pro, *rpoC* Ile491Val). These results suggest that the model can flag previously unclassified variants for closer scrutiny, guiding experimental and epidemiological studies aimed at clarifying their roles in resistance and transmissibility.

Nonetheless, this proof-of-concept study was based on a relatively limited dataset of 249 SNP-specific phylogenetic trees, each comprising 500 samples (leaves). Expanding to larger cohorts with greater genetic diversity is expected to strengthen predictive power and improve generalisation. Overall, this is among the first studies to adapt a phylogenetic tree into a GNN data format for predicting positive selection, demonstrating the potential of aligning data modality with model architecture. An important direction for future work will be adapting the architecture for graph-agnostic learning, allowing the model to transfer knowledge across diverse phylogenetic contexts and potentially detect adaptive signals in a wider range of pathogens. Applied to *M. tuberculosis*, such approaches could yield deeper insights into the evolutionary dynamics of mutations that enhance bacterial survival, ultimately informing strategies for improved disease control and treatment.

**Acknowledgements**

LW is funded by a BBSRC LIDO studentship (Ref. BB/T008709/1). TGC and SC are funded by the UKRI (BBSRC BB/X018156/1; MRC MR/R020973/1, MRC MR/X005895/1; EPSRC EP/Y018842/1)). The funders had no role in the study design, data collection and analysis, the decision to publish, or the preparation of the manuscript. The authors declare that they have no conflicts of interest.

**Author contributions**

LW and JEP conceived and directed the project. LW developed the models under the supervision of SC, TGC and JEP. LW wrote the first draft of the manuscript. All authors commented on and edited various versions of the draft manuscript and approved the final manuscript. LW, TGC and JEP compiled the final manuscript.

**Competing interests**

No potential conflict of interest was reported by the authors.

**Data & Code availability**

The ENA accession code of the samples and the processed data (vcf, tree) are available in the authors' GitHub repository. The complete codebase for preprocessing, model training, and evaluation is also provided. The repository ensures full reproducibility of the results presented in this study and can be accessed at: https://github.com/linfeng-wang/Phylogeny_guided_GNN-new

**Figures and tables**

**Figure 1. Phylogeny-informed graph construction and node feature propagation using attention-based graph neural networks.**

**a)** Simulated phylogenetic tree of eight samples, internal nodes coloured green, mutations presence coloured red. **b)** corresponding pairwise phylogenetic distance (internal node count) matrix, where entries denote branch-length distances; blue cells represent closely related clades selected for subgraph construction after >4 inner node distance cut-off. **c)** Edge weights correspond to phylogenetic distances. Attention is computed using both node features (SNP presence) and edge lengths (branch distances). **d)** The sigma function (σ), in the update equation represents the non-linear activation function (LeakyRelu) applied after neighbour aggregation as a part of the feature extraction steps of the GNN. Attention weights ($\alpha_{ij}$) are computed from both node features and edge length (phylogenetic distances - internal node count), following GAT layer formulation. **d)** Graph Attention Network (GAT) layer update equation with integrated edge-aware bias. Node-based attention is computed using a learnable vector $\alpha$ over the concatenated linear projections of source and target node features, while edge-level influence is incorporated via a scalar product between a learnable parameter $b$ and the edge length $edge_{ij}$. The result is passed through a non-linear activation function $\sigma$, modulating attention weights during message passing.

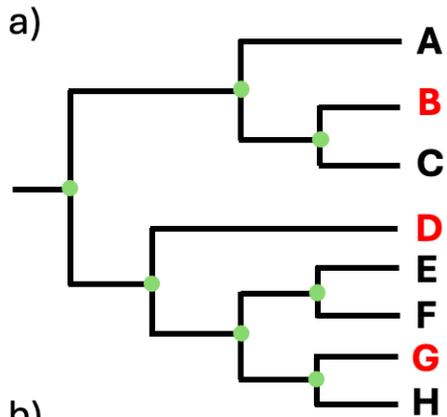
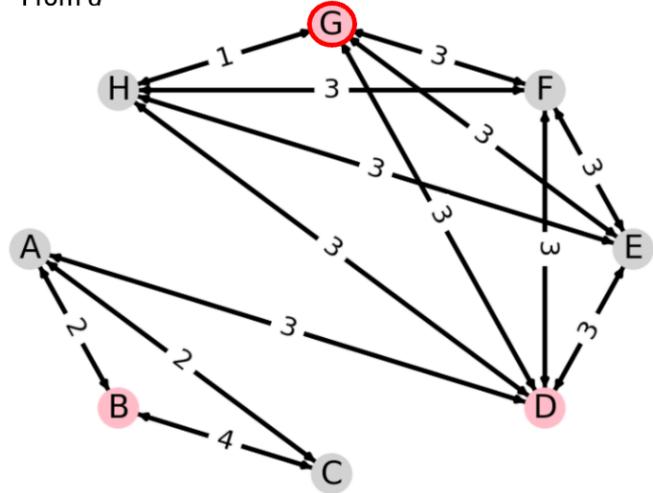

**Figure 2. edge-level attention weights**

Edge-level attention weights in the first GAT layer for a mutation network plot corresponding to the *rpoB* Ser450Leu mutation (red: mutation present, grey: mutation absent). High attention (green/yellow)

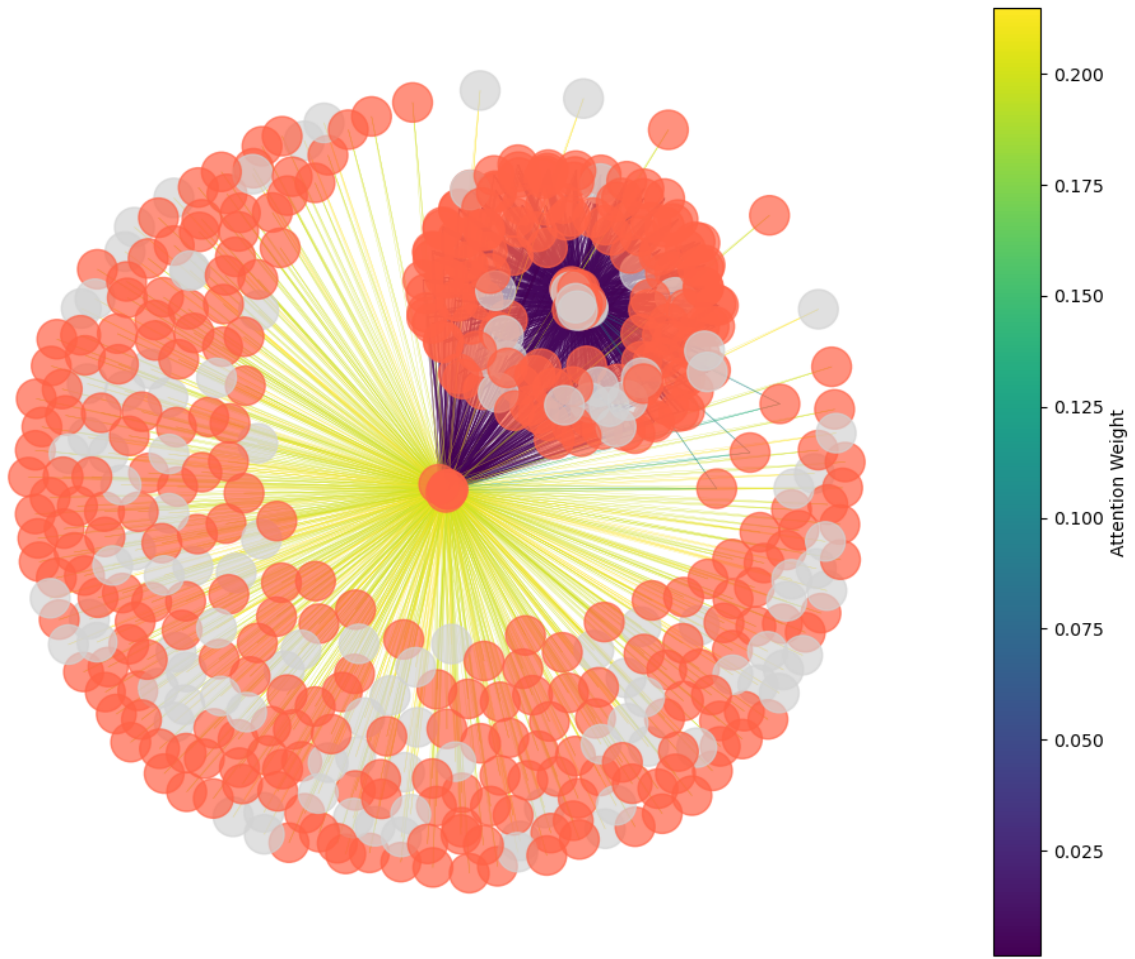

**Table 1. Confusion matrix of GAT model predictions of 50 SNP classifications.**

The matrix summarises the performance of the graph attention network on the test (hold-out) dataset.

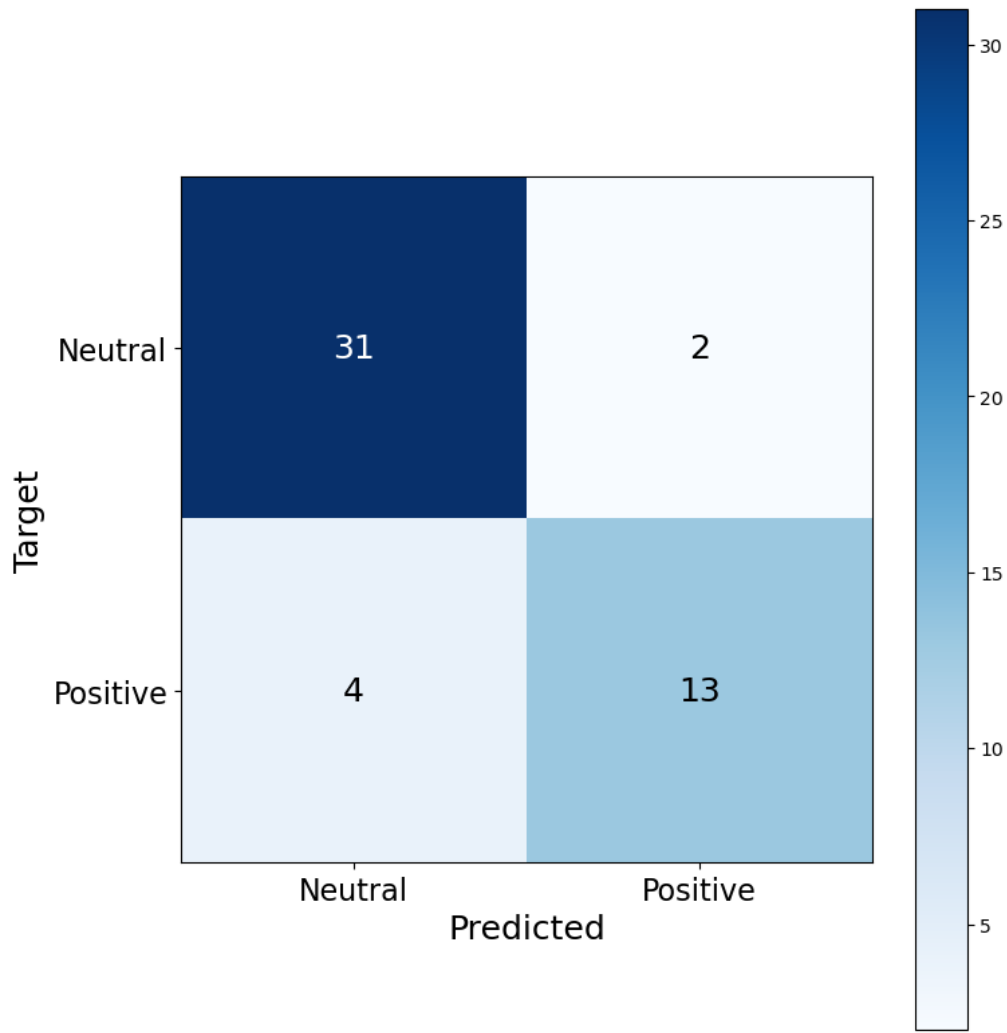

**SUPPLEMENTARY INFORMATION**

**Supplementary Figure S1. SNP specific-Phylogenetic tree instance and its associated GAT (n=500) for the rifampicin resistance**

**(A) Phylogenetic tree for the *rpoB* mutation Ser450Leu** (Mutation presence - red; Mutation absence – grey)

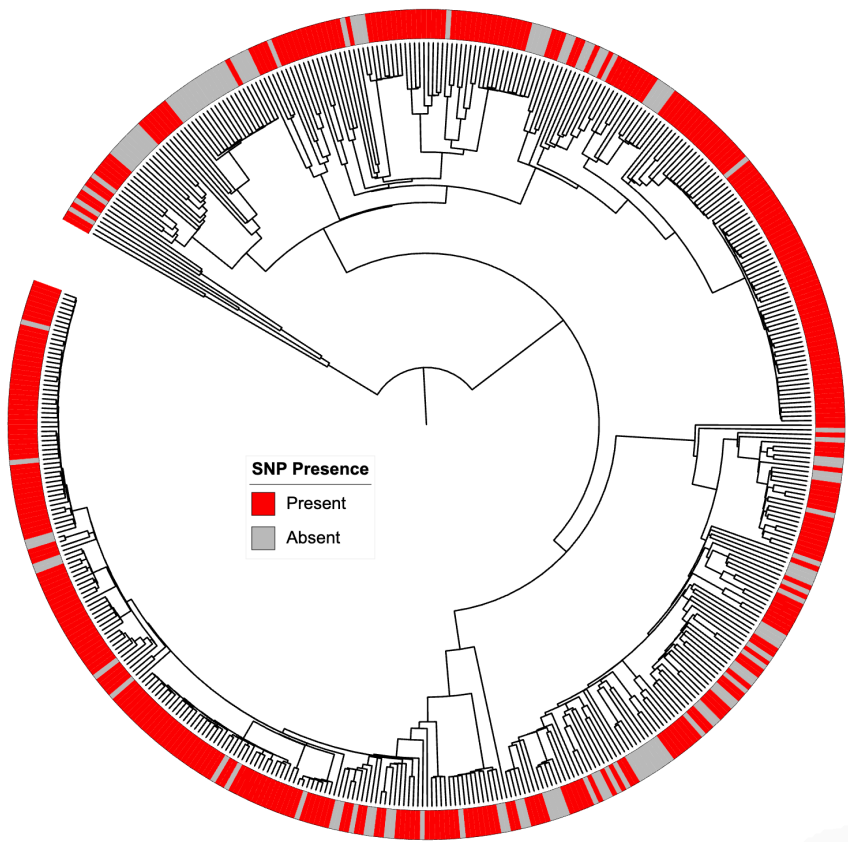

(B) **GAT input graph for (A).** This graph structure (the *rpoB* - Ser450Leu) is used as input for GAT training, allowing the model to propagate and weight information through biologically relevant neighbours. (Mutation presence - red; Mutation absence – grey)

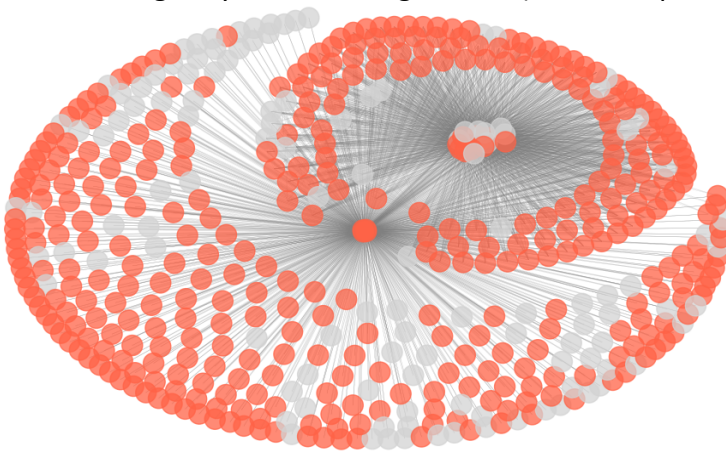

**Table S1. WHO Drug resistance Uncertain variant predictions**

| Gene | Change | Confidence | MDR+ (%) | Arising N | Freq. lineage (%) | Notes |
|---|---|---|---|---|---|---|
| embA | c.-43G>C | 1 | 632 (47.48) | 43 | L2.2.1 (27) | |
| ubiA | Ala249Thr | 1 | 535 (49.95) | 28 | L2.2.2 (88) | |
| rpoC | Val483Gly | 1 | 4513 (49.41) | 37 | L2.2.1 (33) | Compensatory[19] |
| rpoC | Leu516Pro | 0.99 | 521 (48.83) | 35 | L2.2.1 (73) | Compensatory[20] |
| rpoC | Val483Ala | 0.99 | 1570 (49.68) | 33 | L2.2.1 (52) | Putative[21] |
| Rv0010c | c.-80A>G | 0.98 | 64 (26.34) | 30 | L2.2.1 (44) | |
| rpoA | Thr187Ala | 0.98 | 547 (49.64) | 37 | L2.2.1 (85) | Compensatory[22] |
| embA | c.-16C>A | 0.98 | 467 (49.11) | 41 | L2.2.1 (36) | |
| rpoC | Asn698Ser | 0.98 | 408 (49.64) | 29 | L2.2.1 (70) | Compensatory[23] |
| embA | c.-26delA | 0.98 | 32 (50.0) | 16 | L2.2.1 (25) | |
| rpoC | Lys1152Thr | 0.98 | 31 (49.21) | 18 | L2.2.1 (44) | |
| ubiA | Val188Ala | 0.98 | 266 (49.72) | 22 | L4.3.3 (84) | Putative[24] |
| rpoC | Val431Met | 0.98 | 197 (49.0) | 36 | L2.2.1 (43) | |
| rpoC | Asp485Tyr | 0.97 | 705 (49.68) | 32 | L2.2.1 (79) | Compensatory[23] |
| rpoC | Asn416Ser | 0.97 | 264 (48.44) | 34 | L2.2.1 (32) | Compensatory[25] |
| Rv0010c | c.-78A>G | 0.97 | 181 (33.39) | 50 | L2.2.1 (42) | |
| rrl | n.2712C>T | 0.97 | 149 (19.03) | 47 | L2.2.1 (21) | |
| ubiA | Arg240Cys | 0.97 | 135 (48.74) | 28 | L4.6.1.2 (49) | |
| rrs | n.1489C>T | 0.95 | 9 (39.13) | 11 | L2.2.1 (36) | |
| rpoC | Glu1033Lys | 0.95 | 109 (49.1) | 22 | L2.2.1 (46) | |
| rpoC | Ile491Val | 0.93 | 1848 (49.46) | 44 | L2.2.1 (67) | |
| embA | c.-8C>A | 0.88 | 347 (49.36) | 34 | L2.2.1 (81) | |
| Rv1979c | Arg409Gln | 0.87 | 2013 (49.41) | 23 | L2.2.2 (98) | |
| whiB6 | c.-42G>T | 0.8 | 1252 (47.93) | 32 | L2.2.1 (97) | |
| whiB6 | Thr51Pro | 0.66 | 4213 (48.64) | 35 | L2.2.1 (96) | |
| katG | c.-507C>G | 0.62 | 979 (49.57) | 22 | L2.2.1 (98) | |
| rpoC | Lys445Arg | 0.61 | 328 (49.1) | 32 | L2.2.1 (63) | Compensatory[26] |

None of the variants are present in the sensitive samples in the 100K dataset. MDR+ indicates the number of samples containing the variant that fall into multidrug-resistant (MDR), pre-extensively drug-resistant (Pre-XDR), or extensively drug-resistant (XDR) categories. % presence refers to the proportion of MDR+ samples in the 100K database that carry the variant. Freq. lineage denotes the sublineage in which the variant is most observed. % Occurrence indicates the percentage of variant-containing samples that belong to this specific sublineage. Arising N represents the number of distinct sublineages in which the variant is found, capturing the breadth of its phylogenetic distribution.

**Table S2. Model architecture**

| Stage | Layer | Heads | Concat | Output shape |
|---|---|---|---|---|
| Node feats | Input (node feature-SNP presence; Edges-node connection; Edge-length) | – | – | **Node×1** |
| Message pass 1 | GATConv (out=32 per head) | 8 | Yes | **Node×(32x8=256)** |
| Norm/act | BN → ELU → Dropout | – | – | **Node×256** |
| Message pass 2 | GATConv (out=256, head=1) | 1 | No | **Node×256** |
| Residual | BN($x_2$) + $x_1$ → ELU → Dropout | – | – | **Node×256** |
| Readout | GlobalAttention (gate: 256→1) | – | – | **Batch×256** |
| MLP | FC(256→32) → BN → ReLU → Dropout → FC(32→2) | – | – | **Batch×2** |

BN: batch normalisation; ELU: Exponential Linear Unit (ELU) function; FC: Fully connected linear layer; ReLu: Rectified Linear unit